\documentstyle[11pt]{article}
\newcommand{\cl}{\centerline}
\renewcommand{\theequation}{\thesection.\arbic{equation}}

\newcommand\beq{\begin{equation}}
\newcommand\eeq{\end{equation}}
\newcommand\bea{\begin{eqnarray}}
\newcommand\eea{\end{eqnarray}}

\begin{document}

\begin{titlepage}
\setlength{\textwidth}{6.5in}
\setlength{\textheight}{8.5in}
\setlength{\parskip}{0.0in}
\setlength{\baselineskip}{18.2pt}
\cl{\Large{{\bf Sum rules for strange form factors}}}\par
\cl{\Large{{\bf and flavor singlet axial charges}}}\par
\vskip 0.8cm
\cl{Soon-Tae Hong$^{*}$}\par
\vskip 0.4cm
\begin{center}
{Department of Physics and Basic Science Research Institute}\par
{Sogang University, C.P.O. Box 1142, Seoul 100-611, Korea}\par
\end{center}
\vskip 0.3cm
\cl{November 28, 2001}
\vskip 0.5cm
\vfill
\begin{center}
{\bf ABSTRACT}
\end{center}
\begin{quotation}
In chiral models with SU(3) group structure, strange form factors 
of baryon octet are evaluated by constructing their sum rules to yield 
theoretical predictions comparable to the recent experimental data of 
SAMPLE Collaboration.  We also study sum rules for the flavor singlet axial 
currents for the EMC experiment in a modified quark model.  

\vskip 0.5cm
\noindent
---------------------------------------------------------------------\\
\noindent
$^{*}$sthong@ccs.sogang.ac.kr\\
\noindent
\end{quotation}
\end{titlepage}

\section{Introduction}
\setcounter{equation}{0}
\renewcommand{\theequation}{\arabic{section}.\arabic{equation}}

There have been many interesting developments concerning the strange flavor 
structures in the nucleon and the hyperons.  Especially, the internal 
structure of the nucleon is still a subject of great interest to both
experimentalists and theorists.  In 1933, Frisch and 
Stern~\cite{stern33} performed the first measurement of the magnetic moment of 
the proton and obtained the earliest experimental evidence for the internal 
structure of the nucleon.  However, it wasn't until 40 years later that the 
quark structure of the nucleon was directly observed in deep inelastic 
electron scattering experiments and we still lack a quantitative theoretical 
understanding of these properties including the magnetic moments.

Quite recently, the SAMPLE Collaboration~\cite{sample01} reported the
experimental data of the proton strange form factor through parity violating
electron scattering~\cite{mck89}.  To be more precise, they measured the 
neutral weak form factors at a small momentum transfer $Q_{S}^2 = 0.1~{\rm 
(GeV/c)}^2$ to yield the proton strange magnetic form factor in units of 
Bohr nuclear magnetons (n.m.) $G_{M}^{s} (Q_{S}^2)=+0.14 \pm 0.29~{\rm (stat)} 
\pm 0.31~{\rm (sys)}$ n.m.~\cite{sample01}.  This positive experimental 
value is contrary to the negative values of the 
proton strange form factor which result from most of the model calculations 
except the predictions~\cite{hong93,hong97} based on the SU(3) chiral bag 
model~\cite{gerry791} and the recent predictions of the chiral quark soliton model~\cite{kim98} and the heavy baryon chiral 
perturbation theory~\cite{meissner00}.  (See Ref.~\cite{hongpr01} for more 
details.)  

On the other hand, the EMC experiment~\cite{ashman} also reported the highly nontrivial 
data that less than 30\% of the proton spin may be carried by the quark
spin, which is quite different from the well-known prediction from constituent 
quark model.  To explain this discrepancy, it has been proposed~\cite{ansel95} 
that the experimentally measured quantity is not merely the quark spin 
polarization $\Delta\Sigma$ but rather the flavor singlet axial current 
(FSAC) via the axial anomaly mechanism~\cite{treiman96}.  Recently, at the 
quark model renormalization scale, the gluon polarization contribution to the 
FSAC in the chiral bag model has been calculated~\cite{rho00} to yield a 
significant reduction in the relative fraction of the proton spin carried by 
the quark spin, consistent with the small FSAC measured in the EMC experiments. 

In this paper, in the chiral models with SU(3) group structure, we will 
investigate the strange form factors of octet baryons in terms of the sum 
rules of the baryon octet magnetic moments to predict the proton strange form 
factor.  We will also study the modified quark model with SU(3) group 
structure to obtain sum rules for the strange flavor singlet axial current of 
the nucleon in terms of the octet magnetic moments $\mu_{B}$ and the nucleon 
axial vector coupling constant $g_{A}$.  In section 2, we construct the sum 
rules of the baryon octet magnetic moments in the SU(3) chiral models.  In 
section 3 we construct the sum rules for the nucleon strange flavor singlet 
axial current in the modified quark model.  

\section{Strange form factors}
\setcounter{equation}{0}
\renewcommand{\theequation}{\arabic{section}.\arabic{equation}}

Now we consider the baryon magnetic moments in the chiral models such as 
Skyrmion~\cite{skyrme61}, MIT bag~\cite{chodos74} and chiral 
bag~\cite{gerry791} with the general chiral SU(3) group structure.  In 
these models, the EM currents yield the magnetic moment
operators $\hat{\mu}^{i}=\hat{\mu}^{i(3)}+\frac{1}{\sqrt{3}}\hat{\mu}^{i(8)}$
where $\hat{\mu}^{i(a)}=\hat{\mu}^{i(a)}_{CS}+\hat{\mu}^{i(a)}_{FSB}$ with
\begin{eqnarray}
\hat{\mu}^{i(a)}_{CS}&=&-{\cal N}D_{ai}^{8}-{\cal N}^{%
\prime}d_{ipq}D_{ap}^{8} \hat{T}_{q}^{R}+\frac{N_{c}}{2\sqrt{3}}{\cal M}%
D_{a8}^{8}\hat{J}_{i}  \nonumber \\
\hat{\mu}^{i(a)}_{FSB}&=&-{\cal P}D_{ai}^{8}(1-D_{88}^{8})+{\cal Q} \frac{%
\sqrt{3}}{2}d_{ipq}D_{ap}^{8}D_{8q}^{8}
\label{magop}
\end{eqnarray}
where ${\cal M}$, ${\cal N}$, ${\cal N}^{\prime}$, ${\cal P}$ and ${\cal Q}$
are the inertia parameters calculable in the chiral models~\cite{hong97}.

In the higher dimensional irreducible representation of SU(3) group, the 
baryon wave function is described as~\cite{hong93,kim89}
\beq
|B\rangle = |B\rangle_{8}-C_{\bar{10}}^{B}|B\rangle_{\bar{10}}
               -C_{27}^{B}|B\rangle_{27}
\eeq
where the representation mixing coefficients are given by
$C_{\lambda}^{B}=_{\lambda}\langle B|H_{SB}|B\rangle_{8}/(E_{\lambda}-E_{8})$.  
Here $E_{\lambda}$ is the eigenvalue of the eigen equation 
$H_{0}|B\rangle_{\lambda}=E_{\lambda}|B\rangle_{\lambda}$. 
(For explicit expressions for the Hamiltonian $H=H_{0}+H_{SB}$ in the chiral 
models, see Ref.~\cite{hong97}.)  Using the above baryon wave function, 
the spectrum of the magnetic moment operator $\hat{\mu}^{i}$ 
in Eq. (\ref{magop}) has the hyperfine structure 
\begin{eqnarray}
\mu_{p}&=&\frac{1}{10}{\cal M}+\frac{4}{15}({\cal N}+\frac{1}{2}
{\cal N}^{\prime})+\frac{8}{45}{\cal P}-\frac{2}{45}{\cal Q} 
\nonumber \\
& &+m{\cal I}_{2} (\frac{2}{125}{\cal M}+\frac{8}{1125}{\cal N}-\frac{16}{1125}{\cal N}^{\prime}), 
\nonumber \\
\mu_{n}&=&\frac{1}{20}{\cal M}-\frac{1}{5}({\cal N}+\frac{1}{2}
{\cal N}^{\prime})-\frac{1}{9}{\cal P}+\frac{7}{90}{\cal Q} 
\nonumber \\
& &+m{\cal I}_{2} (\frac{31}{750}{\cal M}-\frac{46}{1125}{\cal N}
+\frac{42}{1125}{\cal N}^{\prime}),
\nonumber\\
\mu_{\Lambda}&=&\frac{1}{40}{\cal M}-\frac{1}{10}({\cal N}+\frac12
{\cal N}^\prime)-\frac{1}{10}{\cal P}-\frac{1}{20}{\cal Q}\nonumber\\
& &+m{\cal I}_2
(\frac{9}{500}{\cal M}+\frac{1}{125}({\cal N}-2{\cal N}^\prime )),
\nonumber\\
\mu_{\Xi^0}&=&\frac{1}{20}{\cal M}-\frac{1}{5}({\cal N}+\frac12
{\cal N}^\prime)-\frac{11}{45}{\cal P}-\frac{1}{45}{\cal Q}\nonumber\\
& &+m{\cal I}_2
(\frac{1}{125}{\cal M}+\frac{4}{1125}({\cal N}-2{\cal N}^\prime )),
\nonumber\\
\mu_{\Xi^-}&=&-\frac{3}{20}{\cal M}-\frac{1}{15}({\cal N}+\frac12
{\cal N}^\prime)-\frac{4}{45}{\cal P}-\frac{2}{45}{\cal Q}\nonumber\\
& &+m{\cal I}_2
(\frac{2}{125}{\cal M}+\frac{8}{1125}({\cal N}-2{\cal N}^\prime)),
\nonumber\\
\mu_{\Sigma^+}&=&\frac{1}{10}{\cal M}+\frac{4}{15}({\cal N}+\frac12
{\cal N}^\prime)+\frac{13}{45}{\cal P}-\frac{1}{45}{\cal Q}\nonumber\\
& &+m{\cal I}_2
(\frac{1}{125}{\cal M}+\frac{4}{1125}({\cal N}-2{\cal N}^\prime )),
\nonumber\\
\mu_{\Sigma^0}&=&-\frac{1}{40}{\cal M}+\frac{1}{10}({\cal N}+\frac12
{\cal N}^\prime)+\frac{11}{90}{\cal P}+\frac{1}{36}{\cal Q}
\nonumber\\
& &+m{\cal I}_2
(\frac{37}{1500}{\cal M}-\frac{7}{375}({\cal N}-\frac{17}{21}{\cal N}^\prime)),
\nonumber\\
\mu_{\Sigma^-}&=&-\frac{3}{20}{\cal M}-\frac{1}{15}({\cal N}+\frac12
{\cal N}^\prime)-\frac{2}{45}{\cal P}+\frac{7}{90}{\cal Q}.
\nonumber\\
& &+m{\cal I}_2
(\frac{31}{750}{\cal M}-\frac{46}{1125}({\cal N}-\frac{21}{23}
{\cal N}^\prime)),
\label{octet}
\end{eqnarray}
where the coefficients are solely given by the SU(3) group structure of the chiral 
models and the physical informations such as decay constants and masses are 
included in the above inertia parameters, such as ${\cal M}$, ${\cal N}$ and 
so on.  Note that the SU(3) group structure in the coefficients is generic 
property shared by the chiral models which exploit the hedgehog ansatz solution 
corresponding to the little group SU(2)$\times {\bf Z}_{2}$~\cite{jenkins94}.  
In the chiral perturbation theory to which the hedgehog ansatz does not apply, one can 
thus see the coefficients different from those in Eq. (\ref{octet}) even though 
the SU(3) flavor group is used in the theory~\cite{meissner97}.

Now it seems appropriate to comment on the $1/N_{c}$ 
expansion~\cite{thooft74,witten79,jenkins94,jenkins98}.  In the above relations 
(\ref{octet}), the inertia parameters ${\cal N}$, ${\cal N}^{\prime}$, ${\cal P}$
and ${\cal Q}$ are of order $N_{c}$ while ${\cal M}$ is of order $N_{c}^{-1}$.
However, since the inertia parameter ${\cal M}$ is multiplied by an explicit 
factor $N_{c}$ in Eq. (\ref{magop}), the terms with ${\cal M}$ are of order 
$N_{c}^{0}$.  Moreover, the inertia parameter $m$ is of order of $m_{s}$.  (For 
details of further $1/N_{c}$ and $m_{s}$ orders, see the 
Refs.~\cite{jenkins94,jenkins98}.)

Using the V-spin symmetry sum 
rules~\cite{hong97}, one can obtain the relation
\beq
\frac{1}{2}{\cal M}=\mu_{p}-\mu_{\Xi^{-}}-\frac{1}{3}(\mu_{\Sigma^{+}}
-\mu_{\Xi^{0}})+\frac{4}{3}(\mu_{n}-\mu_{\Sigma^{-}})
\label{m}
\eeq
which will be used later to obtain sum rules of the strange form factors of 
octet baryons.

Now we consider the form factors of the octet baryons which, in the chiral 
models, are definitely extended objects with internal structure associated 
with the electromagnetic (EM) current, to which the photon couples, 
\beq
\hat{V}_{\gamma}^{\mu}=\frac{2}{3}\bar{u}\gamma^{\mu}u
 -\frac{1}{3}\bar{d}\gamma^{\mu}d-\frac{1}{3}\bar{s}\gamma^{\mu}s.
\label{emcurrent}
\eeq
According to the Feynman rules the matrix element of $\hat{V}_{\gamma}^{\mu}$ 
for the baryon with transition from momentum state $p$ to momentum state 
$p+q$ is given by the following covariant decomposition
\beq
\langle p+q|\hat{V}_{\gamma}^{\mu}|p\rangle = 
\bar{u}(p+q) \left[ F_{1B}^{\gamma}(q^{2})\gamma^{\mu}
+\frac{i}{2M_{B}}F_{2B}^{\gamma}(q^{2})\sigma^{\mu\nu}q_{\nu}\right]u(p)
\label{extpf}
\eeq
where $u(p)$ is the spinor for the baryon states and $q$ is the momentum 
transfer and $\sigma^{\mu\nu}=\frac{i}{2}(\gamma^{\mu}\gamma^{\nu}
-\gamma^{\nu}\gamma^{\mu})$ and $M_{B}$ is the baryon mass and 
$F_{1}^{\gamma}$ and $F_{2}^{\gamma}$ are the Dirac and Pauli EM form factors, 
which are Lorentz scalars and $p^{2}=(p+q)^{2}=M_{B}^{2}$ on shell so that 
they depend only on the Lorentz scalar variable $q^{2}(=-Q^{2})$.  We will 
also use the Sachs form factors, which are linear combinations of the Dirac 
and Pauli form factors 
\beq
G_{E}^{\gamma}=F_{1B}^{\gamma}+\frac{q^{2}}{4M_{B}^{2}} F_{2B}^{\gamma},~~~
G_{M}^{\gamma}=F_{1B}^{\gamma}+F_{2B}^{\gamma}
\label{sachs}
\eeq
which can be rewritten as 
\begin{equation}
G_{E,M}^{\gamma}=\frac{2}{3}G_{E,M}^{u}-\frac{1}{3}G_{E,M}^{d}
-\frac{1}{3}G_{E,M}^{s}.
\label{gegamma}
\end{equation}

On the other hand, the neutral weak current operator is given by an expression 
analogous to Eq. (\ref{emcurrent}) but with different coefficients:
\bea
\hat{V}^\mu_Z &=& (\frac{1}{4}-\frac{2}{3}\sin^2 \theta_W)\bar u\gamma^\mu u
        +(-\frac{1}{4}+\frac{1}{3}\sin^2 \theta_W)\bar d\gamma^\mu d\nonumber\\
   & &+(-\frac{1}{4}+\frac{1}{3}\sin^2 \theta_W) \bar s \gamma^\mu s.
\eea
Here the coefficients depend on the weak mixing angle, which has recently
been determined \cite{pdg} with high precision: $\sin^2 \theta_W = 0.2315
\pm 0.0004\>.$  In direct analogy to Eq. (\ref{gegamma}), we have expressions 
for the neutral weak form factors $G_{E,M}^{Z}$ in terms of the different quark 
flavor components:
\bea
G_{E,M}^{Z} &=& (\frac{1}{4}-\frac{2}{3}\sin^2 \theta_W) G_{E,M}^u
 +(-\frac{1}{4}+\frac{1}{3}\sin^2 \theta_W)  G_{E,M}^d \nonumber\\
   & &+(- \frac{1}{4}+\frac{1}{3}\sin^2 \theta_W)  G_{E,M}^s.
\label{gemz}
\eea
Here one notes that the form factors $G_{E,M}^{f}$ $(f=u$, $d$, $s)$ 
appearing in this expression are exactly the same as those in the EM form 
factors, as in Eq. (\ref{gegamma}).

Utilizing isospin symmetry, one then can eliminate the up and down quark 
contributions to the neutral weak form factors by using the proton and neutron
EM form factors and obtain the expressions
\begin{equation}
G_{E,M}^{Z,p} = (\frac{1}{4}-\sin^2 \theta_W) G_{E,M}^{\gamma,p}
                -\frac{1}{4}G_{E,M}^{\gamma,n}-\frac{1}{4}G_{E,M}^{s}.
\label{gemzp}
\end{equation}
It shows how the neutral weak form factors are related to the EM form 
factors plus a contribution from the strange (electric or magnetic) form 
factor.  Measurement of the neutral weak form factor will thus allow (after 
combination with the EM form factors) determination of the strange form factor
of interest. It should be mentioned that there are electroweak radiative
corrections to the coefficients in Eq. (\ref{gemz}), which are generally 
small corrections, of order 1-2\%, and can be reliably 
calculated~\cite{musolf90}.

The EM form factors present in Eq. (\ref{gemzp}) are very accurately known 
(1-2 \%) for the proton in the momentum transfer region $Q^2 < 1$ 
(GeV/c)${}^2$. The neutron form factors are not known as accurately as the 
proton form factors (the electric form factor $G_E^n$ is at present rather 
poorly constrained by experiment), although considerable work to improve our 
knowledge of these quantities is in progress.  Thus, the present lack of 
knowledge of the neutron form factors will significantly hinder the 
interpretation of the neutral weak form factors.

At zero momentum transfer, one can have the relations between the EM form 
factors and the static physical quantities of the baryon octet, namely 
$G_{E}^{\gamma}(0)=Q_{B}$ and $G_{M}^{\gamma}(0)=\mu_{B}$ with the electric 
charge $Q_{B}$ and magnetic moment $\mu_{B}$ of the baryon.  In the strange 
flavor sector, the Sachs magnetic form factor in Eq. (\ref{sachs}) yields the 
strange form factors of baryon octet degenerate in isomultiplets 
$F_{2B}^{s}(0) = G_{M}^{s}(0)-F_{1B}^{s}(0)$ where $F_{1B}^{s}=-3Q_{B}^{s}$ 
with the fractional strange EM charge $Q_{B}^{s}$.  Here note that one can 
express the slope of $G_{E}^{s}$ at $Q^2=0$ in the usual fashion in terms of 
a strangeness radius $r_{s}$ defined as 
$r^{2}_{s}=-6\left[dG_{E}^{s}/dQ^2\right]_{Q^2=0}$.

Now we construct model independent sum rules for the strange form factors of 
baryon octet in the chiral models with the SU(3) flavor group structure.  
Since the nucleon has no net strangeness the nucleon strange form factor is 
given by~\cite{hong97}
\bea
F_{2N}^{s}(0)&=&\frac{7}{20}{\cal M}-\frac{1}{15}({\cal N}+\frac{1}{2}
{\cal N}^{\prime})-\frac{1}{15}{\cal P}-\frac{1}{30}{\cal Q}\nonumber \\
& &+m{\cal I}_{2} (-\frac{43}{750}{\cal M}+\frac{38}{1125}{\cal N}
-\frac{26}{1125}{\cal N}^{\prime}). 
\label{psff}
\eea
Substituting Eq. (\ref{m}) into the relation $F_{2N}^{s}(0)+\mu_{p}+\mu_{n}
-\frac{1}{2}{\cal M}=0$ obtained from Eqs. (\ref{octet}) and (\ref{psff}), we 
obtain the sum rules for the nucleon strange form factor
\beq
F_{2N}^{s}(0)=\mu_{p}-\mu_{\Xi^{-}}-(\mu_{p}+\mu_{n})
-\frac{1}{3}(\mu_{\Sigma^{+}}-\mu_{\Xi^{0}})
+\frac{4}{3}(\mu_{n}-\mu_{\Sigma^{-}})
\label{psff2}
\eeq
which, at least within the SU(3) flavor chiral models, is independent of the 
values of the model dependent inertia parameters.  Inserting into 
Eq. (\ref{psff2}) the experimental data for the baryon octet magnetic moments, 
one can evaluate the nucleon strange form factor
\beq
F_{2N}^{s}(0)=G_{M}^{s}(0)=0.32~{\rm n.m.}.
\label{f2nsh}
\eeq

On the other hand, the quantities $G_{E,M}^Z$ in Eq. (\ref{gemzp}) for the 
proton can be determined via elastic parity-violating electron scattering to 
yield the experimental data $G_{M}^{s} (Q_{S}^2) = +0.14 \pm 0.29~{\rm (stat)} 
\pm 0.31~{\rm (sys)}$ n.m.~\cite{sample01} for the proton strange magnetic 
form factor.  Here one notes that the prediction for the proton strange form 
factor (\ref{f2nsh}) obtained from the sum rule (\ref{psff2}) is comparable 
to the SAMPLE data.  Moreover, from the relation (\ref{gemzp}) at zero 
momentum transfer, the neutral weak magnetic moment of the nucleon can be 
written in terms of the nucleon magnetic moments and the proton strange form 
factor~\cite{mckeown002}
\beq
4\mu_{p}^{Z}=\mu_{p}-\mu_{n}-4\sin^{2}\theta_{W}\mu_{p}-F_{2N}^{s}(0).
\label{mupz}
\eeq

Next, we obtain the other octet baryon strange form factors~\cite{hong97}
\bea
F_{2\Lambda}^{s}(0)&=&\frac{9}{20}{\cal M}+\frac{1}{5}({\cal N}+\frac{1}{2}
{\cal N}^{\prime})+\frac{1}{5}{\cal P}+\frac{1}{10}{\cal Q} 
\nonumber \\
& &+m{\cal I}_{2} (-\frac{9}{250}{\cal M}-\frac{2}{125}{\cal N}
+\frac{4}{125}{\cal N}^{\prime})-1,
\nonumber\\
F_{2\Xi}^{s}(0)&=&\frac{3}{5}{\cal M}+\frac{4}{15}({\cal N}+\frac{1}{2}
{\cal N}^{\prime})+\frac{1}{3}{\cal P}+\frac{1}{15}{\cal Q} 
\nonumber \\
& &+m{\cal I}_{2} (-\frac{3}{125}{\cal M}-\frac{4}{375}{\cal N}
+\frac{8}{375}{\cal N}^{\prime})-2,
\nonumber\\
F_{2\Sigma}^{s}(0)&=&\frac{11}{20}{\cal M}-\frac{1}{5}({\cal N}+\frac{1}{2}
{\cal N}^{\prime})-\frac{11}{45}{\cal P}-\frac{1}{18}{\cal Q} 
\nonumber \\
& &+m{\cal I}_{2} (-\frac{37}{750}{\cal M}+\frac{14}{375}{\cal N}
-\frac{34}{1125}{\cal N}^{\prime})-1,
\eea
which, similarly to the nucleon strange form factors, can be rewritten in 
terms of the octet magnetic moments to yield the sum rules for the other 
octet strange form factors 
\bea
F_{2\Lambda}^{s}(0)&=&\mu_{p}-\mu_{\Xi^{-}}-2\mu_{\Lambda}
-\frac{1}{3}(\mu_{\Sigma^{+}}-\mu_{\Xi^{0}})+\frac{4}{3}(\mu_{n}
-\mu_{\Sigma^{-}})-1, \nonumber\\
F_{2\Xi}^{s}(0)&=&\mu_{p}-\mu_{\Xi^{-}}-(\mu_{\Xi^{0}}+\mu_{\Xi^{-}})
-\frac{1}{3}(\mu_{\Sigma^{+}}-\mu_{\Xi^{0}})+\frac{4}{3}
(\mu_{n}-\mu_{\Sigma^{-}})-2,\nonumber\\
F_{2\Sigma}^{s}(0)&=&\mu_{p}-\mu_{\Xi^{-}}-(\mu_{\Sigma^{+}}+\mu_{\Sigma^{-}})
-\frac{1}{3}(\mu_{\Sigma^{+}}-\mu_{\Xi^{0}})+\frac{4}{3}
(\mu_{n}-\mu_{\Sigma^{-}})-1.
\nonumber\\
\label{sumrules7}
\eea
Note that these sum rules (\ref{psff2}) and (\ref{sumrules7}) are extracted 
only from the intrinsic SU(3) flavor group structures of the octet baryons.  
Using the experimental data for the known baryon octet magnetic moments, we 
can predict the octet baryon strange form factors as shown in Table 1.  We 
also evaluate the strange form factors by using the theoretical predictions 
from the chiral bag model, Skyrmion model and chiral quark soliton model as 
input data of the sum rules (\ref{psff2}) and (\ref{sumrules7}) given in the 
SU(3) flavor chiral models.  Here one notes that, since the values of the 
magnetic moments used in the theoretical model predictions of the baryon 
strange form factors have already had discrepancies deviated from the 
corresponding experimental values of the baryon magnetic moments, the 
predicted values of the baryon octet strange form factors listed in Table 1 
are unreliably sensitive in the strange flavor channel.    

\section{Strange flavor singlet axial currents}
\setcounter{equation}{0}
\renewcommand{\theequation}{\arabic{section}.\arabic{equation}}

In this section, we consider a modified quark model~\cite{ls94}.  In the 
nonrelativistic quark model, the quarks possess the static properties such 
as mass, electromagnetic charge and magnetic moments, which are independent 
of their surroundings.  However this assumption seems to be irrelevant to 
the realistic experimental situation.  In the literature~\cite{ls94}, the 
magnetic moments of the quarks were proposed to be different in the different 
isomultiplets, but to be the same within an isomultiplet.  The magnetic moments 
are then given by 
\begin{equation}
\mu_{B}=\mu_{u}^{B}\Delta u^{B}+\mu_{d}^{B}\Delta d^{B}
+\mu_{s}^{B}\Delta s^{B},  \label{mubi}
\end{equation}
where $\mu_{f}^{B}$ is an effective magnetic moment of the quark of flavor $f
$ for the baryon $B$ degenerate in the corresponding baryon isomultiplet, and 
$\Delta f^{B}$ is the spin polarization for the baryon.

Using the SU(3) charge symmetry one can obtain the magnetic moments of
the octet baryons as follows~\cite{ls94}\footnote{In the 
literature~\cite{leinweber01}, the similar equalities are used in connection 
with the quark-loops.} 
\begin{eqnarray}
\mu_{p}&=&\mu_{u}^{N}\Delta u +\mu_{d}^{N}\Delta d +\mu_{s}^{N}\Delta s 
\nonumber \\
\mu_{n}&=&\mu_{u}^{N}\Delta d +\mu_{d}^{N}\Delta u +\mu_{s}^{N}\Delta s 
\nonumber \\
\mu_{\Lambda}&=&\frac{1}{6}(\mu_{u}^{\Lambda}+\mu_{d}^{\Lambda})(\Delta u
+4\Delta d +\Delta s)+\frac{1}{3}\mu_{s}^{\Lambda}(2\Delta u -\Delta d
+2\Delta s)  \nonumber \\
\mu_{\Xi^0}&=&\mu_{u}^{\Xi}\Delta d +\mu_{d}^{\Xi}\Delta s +\mu_{s}^{\Xi}
\Delta u  \nonumber \\
\mu_{\Xi^-}&=&\mu_{u}^{\Xi}\Delta s +\mu_{d}^{\Xi}\Delta d +\mu_{s}^{\Xi}
\Delta u  \nonumber \\
\mu_{\Sigma^+}&=&\mu_{u}^{\Sigma}\Delta u +\mu_{d}^{\Sigma}\Delta s
+\mu_{s}^{\Sigma}\Delta d  \nonumber \\
\mu_{\Sigma^0}&=&\frac{1}{2}(\mu_{u}^{\Sigma}+\mu_{d}^{\Sigma})(\Delta u
+\Delta s)+\mu_{s}^{\Sigma}\Delta d  \nonumber \\
\mu_{\Sigma^-}&=&\mu_{u}^{\Sigma}\Delta s +\mu_{d}^{\Sigma}\Delta u
+\mu_{s}^{\Sigma}\Delta d.  
\label{mus}
\end{eqnarray}
Here one notes that it is difficult to figure out which terms are of the 
order of $m_{s}$ and whether $\Delta f$ contain symmetry breaking or 
whether the symmetry breaking manifests itself only in the fact that 
the quark magnetic moments are different for different baryons. 

After some algebra we obtain the novel sum rules for spin polarizations 
$\Delta f$ with the flavor $f$ in terms of the octet magnetic moments 
$\mu_{B}$ and the nucleon axial vector coupling constant $g_{A}$ 
\begin{eqnarray}
\Delta u &=& g_{A}\frac{R_{\Sigma}-2R_{\Xi}+R_{S}+3R_{\Xi}(R_{S}-R_{\Sigma})}
{3(R_{\Sigma}-R_{\Xi})(1-R_{S})}  \nonumber \\
\Delta d &=& g_{A}\frac{-2R_{\Sigma}+R_{\Xi}+R_{S}+3R_{\Sigma}(R_{S}-R_{\Xi})}
{3(R_{\Sigma}-R_{\Xi})(1-R_{S})}  \nonumber \\
\Delta s &=& g_{A}\frac{R_{\Sigma}+R_{\Xi}-2R_{S}+3(R_{S}^{2}
-R_{\Sigma}R_{\Xi})}{3(R_{\Sigma}-R_{\Xi})(1-R_{S})}
\label{udsdels}
\end{eqnarray}
with 
\begin{eqnarray}
R_{N}&=&\frac{\mu_{p}+\mu_{n}}{\mu_{p}-\mu_{n}}  \nonumber \\
R_{\Sigma}&=&\frac{\mu_{\Sigma^+}+\mu_{\Sigma^-}}{\mu_{\Sigma^+}
-\mu_{\Sigma^-}}  \nonumber \\
R_{\Xi}&=&\frac{\mu_{\Xi^0}+\mu_{\Xi^-}}{\mu_{\Xi^0} -\mu_{\Xi^-}}  
\nonumber\\
R_{S}&=&(R_{N}R_{\Sigma}+R_{\Sigma}R_{\Xi}-R_{\Xi}R_{N})^{1/2}
\end{eqnarray}
where we have assumed the isospin symmetry $\mu_{u}^{B}=-2\mu_{d}^{B}$. Here 
one notes that the above sum rules (\ref{udsdels}) are given only in terms of 
the physical quantities, the coupling constant $g_{A}$ and baryon octet 
magnetic moments $\mu_{B}$, which are independent of details involved in the 
modified quark model, as in the sum rules in Eqs. (\ref{psff2}) and 
(\ref{sumrules7}).  Moreover these sum rules are governed only by the SU(3) 
flavor group structure of the models.

Using the experimental data for $g_{A}$ and $\mu_{B}$, we obtain the strange 
flavor spin polarization $\Delta s$ 
\begin{equation}
\Delta s=-0.20
\label{udspred}
\end{equation}
which is comparable to the recent SMC measurement $\Delta s=-0.12\pm
0.04$~\cite{adams95} and, together with the other flavor spin polarizations 
$\Delta u=0.88$ and $\Delta d=-0.38$, one can arrive at the flavor singlet 
axial current of the nucleon as follows\footnote{In fact, in the literature~\cite{ls94}, 
$\Delta\Sigma$ is evaluated using the sum rule for $\Delta\Sigma$.  However, 
here we have explicitly obtained the sum rules for its flavor components 
$\Delta f$ $(f=u,d,s)$ and $F_{2N}^{s(0)}$ as shown in Eqs. 
(\ref{udsdels}) and (\ref{musdels2}) to predict the values for 
$\Delta s$ and $F_{2N}^{s(0)}$ in Eqs. (\ref{udspred}) and (\ref{f2ns062}).} 
\begin{equation}
\Delta \Sigma =\Delta u+\Delta d +\Delta s = 0.30
\label{delsigma}
\end{equation}
which is comparable to the recent value $\Delta \Sigma=0.28$ obtained from 
the deep inelastic lepton-nucleon scattering experiments~\cite{dis}.
Here note that the strange flavor singlet axial current $\Delta s$ in Eq. 
(\ref{udspred}) is significantly noticeable even though the flavor singlet 
axial current $\Delta \Sigma$ in Eq. (\ref{delsigma}) is not quite large.  
The above predictions are quite consistent with the analysis in the 
literature~\cite{savage97} where $m_{s}{\rm ln}m_{s}$ corrections are used to 
predict $\Delta u=0.77\pm 0.04$, $\Delta d=-0.49\pm 0.04$ and $\Delta s=-0.18\pm 0.09$.

Now it seems appropriate to discuss the strange form factor in this modified 
quark model.  Exploiting the relations (\ref{mus}), together with the isospin 
symmetry $\mu_{u}^{B}=-2\mu_{d}^{B}$, one can easily obtain
\bea
\mu_{p}+\mu_{n}&=&-\mu_{d}^{N}(\Delta u+\Delta d)+2\mu_{s}^{N}\Delta s,
\nonumber\\
\mu_{p}-\mu_{n}&=&-3\mu_{d}^{N}(\Delta u-\Delta d).
\eea
We thus arrive at the sum rule for the nucleon strange form factor in the 
modified quark model
\beq
F_{2N}^{s(0)}=-3\mu_{s}^{N}\Delta s
=-\frac{3}{2}(\mu_{p}+\mu_{n})+\frac{1}{2}(\mu_{p}-\mu_{n})
\frac{\Delta u+\Delta d}{\Delta u-\Delta d}.
\label{musdels2}
\eeq
Substituting the experimental values for $\mu_{p}$ and $\mu_{n}$, and the above 
predictions $\Delta u=0.81$ and $\Delta d=-0.44$, we obtain 
\beq
F_{2N}^{s(0)}=-0.39~{\rm n.m.},
\label{f2ns062}
\eeq
which reveals the discrepancy from the SAMPLE experimental values, differently from 
the prediction (\ref{f2nsh}) of the SU(3) chiral model case.  However, as expected, 
this result is quite comparable to the prediction in the 
literature~\cite{leinweber01} where, similar to Eq. (\ref{mus}), the SU(3) 
charge symmetry relations with the quark-loops are used.  The difference between the 
predictions of $F_{2N}^{s(0)}$ in the SU(3) modified quark model and the SU(3) chiral 
model originates from the assumptions of these models, for instance, those in the 
SU(3) modified quark model that the magnetic moments of the quarks are different in the 
different isomultiplets, but do not change within an isomultiplet.  

On the other hand, in this modified quark model, $\Delta f$ are defined through the 
semileptonic hyperon decays and thus the $\Sigma\rightarrow n$ decay is not well 
reproduced since $g_{A}^{\Sigma n}=\Delta d-\Delta s=-0.18$ is quite different from 
its experimental value $g_{A}^{\Sigma n}=-0.340\pm 0.017$~\cite{pdg}.  Moreover, 
the SU(3) symmetry breaking in the hyperon semileptonic decays can be parameterized 
by the value of the nonsinglet axial charge $a_{8}=\Delta u+\Delta d-2\Delta s$ 
in the hyperon $\beta$-decay~\cite{leader00}.  Exploiting the above values for 
$\Delta f$ in the modified quark model, we obtain the prediction $a_{8}=0.90$, which is 
quite higher than the standard SU(3) value $a_{8}=3F-D=0.579\pm 0.025$~\cite{pdg,leader00}.  
Note that the SU(3) Skyrmion model~\cite{hong92} and large $N_{c}$ QCD~\cite{jenkins98} 
predict $a_{8}=0.41$ and $a_{8}=0.30$, respectively.  It is interesting to see 
that the large value of $a_{8}$ in the modified quark model is incompatible with the 
SAMPLE experimental values. 

\section{Conclusions}
\setcounter{equation}{0}
\renewcommand{\theequation}{\arabic{section}.\arabic{equation}}
\renewcommand{\theequation}{\arabic{equation}}

In summary, we have investigated the strange flavor structure of the octet 
baryon magnetic moments in the chiral models with SU(3) group structure.  
The strange form factors of octet baryons are explicitly constructed in terms 
of the sum rules of the baryon octet magnetic moments to yield the 
theoretical predictions.  Especially in case of using the experimental data 
for the baryon magnetic moments as input data of the sum rules, the 
predicted proton strange form factor is comparable to the recent SAMPLE 
experimental data.

On the other hand, we have studied the modified quark model with SU(3) group 
structure, where the magnetic moments of the quarks are different in the 
different isomultiplets, but do not change within an isomultiplet.  In this 
model, we have obtained the sum rules for the spin polarizations $\Delta f$ 
with the flavor $f$ $(f=u,d,s)$ in terms of the octet magnetic moments 
$\mu_{B}$ and the nucleon axial vector coupling constant $g_{A}$, to yield the 
flavor singlet axial current of the nucleon, comparable to the recent 
experimental data.  Moreover, the strange flavor spin polarization has been shown 
to be quite noticeable.  However, exploiting the sum rule for the nucleon strange 
form factor constructed in the modified quark model, we have obtained the 
prediction, which shows discrepancy from the SAMPLE experimental values but is 
comparable to the prediction in the previous literature.  

\vskip 1.0cm 
STH would like to thank Bob McKeown for helpful discussions and kind concerns at 
Kellogg Radiation Laboratory, Caltech where a part of this work has been done, and 
M. Rho for useful discussions and comments.  He also would like to thank Y.M. Kim, 
K. Kubodera, H.K. Lee, F. Myhrer, Y. Oh and B.Y. Park for helpful discussions, and 
acknowledges financial support in part from the Korea Research Foundation, Grant No. 
KRF-2001-DPP0083.

\newpage
\begin{table}[t]
\caption{The baryon octet strange form factors in units of Bohr nuclear 
magnetons calculated via model independent relations.  For input data for the 
baryon octet magnetic moments we have used the experimental data (Exp) 
and the theoretical predictions from the chiral bag model (CBM), Skyrmion 
model (SM) and chiral quark soliton model (CQSM).}
\begin{center}
\begin{tabular}{lrrrr}
\hline
Input &$F_{2N}^{s}(0)$ &$F_{2\Lambda}^{s}(0)$ &$F_{2\Xi}^{s}(0)$ 
      &$F_{2\Sigma}^{s}(0)$\\ 
\hline
Exp  &0.32   &1.42 &1.10 &$-1.10$\\ 
CBM  &0.30   &0.49 &0.25 &$-1.54$\\ 
SM   &$-0.02$&0.51 &0.09 &$-1.75$\\
CQSM &$-0.02$&1.06 &0.86 &$-1.89$\\ 
\hline\\
\end{tabular}
\end{center}
\end{table}

\begin{thebibliography}{99}

\bibitem{stern33}  R. Frisch and O. Stern, Z. Physik 85, 4 (1933) 4.
\bibitem{sample01} R. Hasty et al., Science 290 (2000) 2117; 
D.T. Spayde et al., Phys. Rev. Lett. 84 (2000) 1106; 
B. Mueller et al., Phys. Rev. Lett. 78 (1997) 3824.  
\bibitem{mck89}  R.D. McKeown, Phys. Lett. B 219 (1989) 140; 
E.J. Beise and R.D. McKeown, Comm. Nucl. Part. Phys. 20 (1991) 105;
R.D. McKeown, {\it New Directions in Quantum 
Chromodynamics} (AIP, Melville, New York 1999) eds. C.R. Ji and D.P. Min.
\bibitem{hong93} S.T. Hong and B.Y. Park, Nucl. Phys. A 561 (1993) 525.
\bibitem{hong97} S.T. Hong, B.Y. Park and D.P. Min, Phys. Lett. B 414 (1997) 229.
\bibitem{gerry791}  G.E. Brown and M. Rho, Phys. Lett. B 82 (1979) 177.
\bibitem{kim98} H.C. Kim, M. Praszalowicz, M.V. Polyakov and K. Goeke, Phys. 
                Rev. D 58 (1998) 114027.
\bibitem{meissner00} Ulf-G. Meissner, Nucl. Phys. A 666-A 667 (2000) 51; 
C.M. Maekawa and U. van Kolck, Phys. Lett. B 478 (2000) 73.
\bibitem{hongpr01} S.T. Hong and Y.J. Park, Phys. Rep. 358 (2002) 143.
\bibitem{ashman} J. Ashman et al., Nucl. Phys. B 238 (1990) 1; 
Phys. Lett. B 206 (1988) 364.
\bibitem{ansel95} M. Anselmino, A. Efremov and E. Leader, Phys. Rep. 261 (1995) 1.
\bibitem{treiman96} S.B. Treiman, R. W. Jackiw, B. Zumino and E. Witten, {\it 
Current Algebra and Anomalies} (World Scientific, Singapore 1996).
\bibitem{rho00} H.J. Lee, D.P. Min, B.Y. Park, M. Rho and V. Vento, Phys. Lett. 
B 491 (2000) 257.
\bibitem{skyrme61} T.H.R. Skyrme, Proc. Roy. Soc. A 260 (1961) 127; 
G. S. Adkins, C. R. Nappi and E. Witten, Nucl. Phys. B 228 (1983) 552; 
I. Zahed and G.E. Brown, Phys. Rep. 142 (1986) 1.
\bibitem{chodos74} A. Chodos et al., Phys. Rev. D 9 (1974) 3471.
\bibitem{kim89} H. Yabu and K. Ando, Nucl. Phys. B 301 (1988) 601; 
                J.H. Kim, C.H. Lee and H.K. Lee, Nucl. Phys. A 501 (1989) 835;
                H.K. Lee and D.P. Min, Phys. Lett. B 219 (1989) 1;  
                S.T. Hong and G.E. Brown, Nucl. Phys. A 564 (1993) 491.
\bibitem{jenkins94} R. Dashen, E. Jenkins and A.V. Manohar, Phys. Rev. D 49 (1994) 4713; 
R. Dashen, E. Jenkins and A.V. Manohar, Phys. Rev. D 51 (1995) 3697; J. Dai, R. Dashen, 
E. Jenkins and A.V. Manohar, Phys. Rev. D 53 (1996) 273.
\bibitem{meissner97} Ulf-G. Meissner and S. Steininger, Nucl. Phys. B 499 (1997) 349. 
\bibitem{thooft74} G. t'Hooft, Nucl. Phys. B 72 (1974) 461. 
\bibitem{witten79} E. Witten, Nucl. Phys. B 160 (1979) 57.
\bibitem{jenkins98} R. Flores-Mendieta, E. Jenkins and A.V. Manohar, Phys. Rev. D 58 
(1998) 94028.
\bibitem{pdg} Particle Data Group, Eur. Phys. J. C 15 (2000) 1.
\bibitem{musolf90} M.J. Musolf and B.R. Holstein, Phys. Lett. B 242 (1990) 461; 
M.J. Musolf, T.W. Donnelly, J. Dubach, S.J. Pollock, S. Kowalski and E.J. Beise, 
Phys. Rep. 239 (1994) 1.
\bibitem{mckeown002} R.D. McKeown, Prog. Part. Nucl. Phys. 44 (2000) 313.
\bibitem{ls94} J. Linde and H. Snellman, Z. Phys. C 64 (1994) 73.
\bibitem{leinweber01} D.B. Leinweber and A.W. Thomas, Nucl. Phys. A 684 (2001) 35.
\bibitem{adams95} D. Adams et al., Phys. Lett. B 357 (1995) 248.
\bibitem{dis} G.K. Mallot, {\it Proceedings of the 12th International Symposium 
on High-Energy Spin Physics, Amsterdam 1996} (World Scientific, Singapore 
1997) eds. C.W. de Jager et al.. 
\bibitem{savage97} M.J. Savage and J. Walden, Phys. Rev. D 55 (1997) 5376.
\bibitem{leader00} E. Leader, A.V. Sidorov and D.B. Stamenov, Phys. Lett. B 488 (2000) 283.
\bibitem{hong92} S.T. Hong and B.Y. Park, Int. J. Mod. Phys. E 1 (1992) 131. 
\end{thebibliography}
\end{document}